\documentclass[aps,prl,print,groupedaddress,twocolumn,showpacs]{revtex4-1}

\usepackage{graphicx}
\usepackage{dcolumn}
\usepackage{bm}

\begin{document}

\title{Interplay between the coherent and incoherent transport in quantum Hall bilayers}

\author{Ding Zhang}
\author{Xuting Huang}
\author{Werner Dietsche}
\author{Maik Hauser}
\author{Klaus von Klitzing}
\affiliation{ Max Planck Institute for Solid State Research, Heisenbergstrasse 1, 70569 Stuttgart, Germany}

\date{\today}

\begin{abstract}
We systematically study the coherent transport (Josephson tunneling and counterflow current) and its breakdown which leads to incoherent charge flow in in the excitonic BCS condensate formed in GaAs bilayers at $\nu_{tot}=1/2+1/2$. The Josephson currents in samples with three different interlayer distances vary by four orders of magnitude. In contrast, the breakdown thresholds for the $\nu_{tot}=1$ quantum Hall state are comparable. Furthermore, Coulomb drag in a Corbino ring reveals that the coherent counterflow current coexists with the dissipative charge transport.
\end{abstract}


\maketitle

Quantum coherence is the key ingredient of many exotic states of matter. Epitomes are the superconductivity and the Bose-Einstein condensation. Interestingly, much of the physics governing the aforementioned states has been recently encountered in a quantum Hall system~\cite{Yoshioka1989,Fer1989,Wen1992,Girvin2003}. In a double-quantum well heterostructure, interlayer coherence emerges at a total Landau level filling of one, i.e. $\nu_{tot}=1$~\cite{Spi2000,Kell2004,Tut2004,Wie2004,Tie2008,Huang2012,Nandi2012}. This spontaneous phase coherence renders the filled and vacant electronic states in the two layers into a BCS-condensate of interlayer excitons.

Experimental evidence gives strong support for the scenario of an exciton condensate. Interlayer tunneling experiments have revealed that a large dc current can be passed from one layer to the other without building up any significant voltage drop~\cite{Spi2000,Tie2008}. Due to its resemblance to a superconductor junction, the tunneling current is therefore considered to be the interlayer Josephson current. Further developments using Corbino samples even allowed adjusting the Josephson current via the bulk exciton condensate~\cite{Huang2012}. The Josephson current induced on one edge shifts accordingly with a second Josephson current injected on the opposite edge. By exploiting the unique topology of a Corbino configuration, it has been established that sending charges across the annulus in one layer drags charges dissipationless in the opposite layer ~\cite{Tie2008}, even in the absence of any Josephson coupling~\cite{Nandi2012}. This remarkable effect attests to the existence of a coherent counterflow current~\cite{Su2008}.

Here,  we enhance the coherent effects in the bilayer system by consecutively reducing the interlayer distance
and study systematically the influence of interlayer distance on both the coherent processes as well as the incoherent transport. We find that the maximum Josephson current changes rapidly with the interlayer distance whereas the thresholds for the incoherent intralayer process are comparable among all bilayers investigated. Coulomb drag experiments in Corbino samples  demonstrate that these incoherent intralayer processes can coexist with a coherent counter-flow current.

The wafers were grown by molecular-beam epitaxy and contain two 19 nm wide GaAs quantum wells in which the electrons reside. Symmetric doping with spacer thicknesses of about 200 nm on each side
results in rather similar electron densities $4\times 10^{10}$~cm$^{-2}$ and mobilities $5\times 10^5$~cm$^2$/V s for the individual layers. The quantum wells are separated by alternating layers of AlAs (1.7~nm) and GaAs (0.28~nm). The three wafers contain four, five, or six AlAs/GaAs periods corresponding to about 8, 10 or 12 nm barrier thicknesses ($b$), respectively.  Separate contacts to the two quantum wells are achieved by the selective depletion technique~\cite{Eis1990,Rub1998}. 
Balanced electron densities of about $2\times 10^{10}$~cm$^{-2}$ are adjusted using front and back gates. 
At these densities the  $\nu_{tot}=1$ state is reached at perpendicular magnetic fields $B$ of about 2 Tesla where the $d/l_B$ ratio becomes less than about 2. Here $d$ is the average distance between the 2DEG layers and $l_B={\sqrt{\hbar/eB}}$ is the magnetic length. 
All the measurements were carried out in a dilution refrigerator at base temperatures below 20 mK.

\begin{figure}
 \includegraphics[width=86mm]{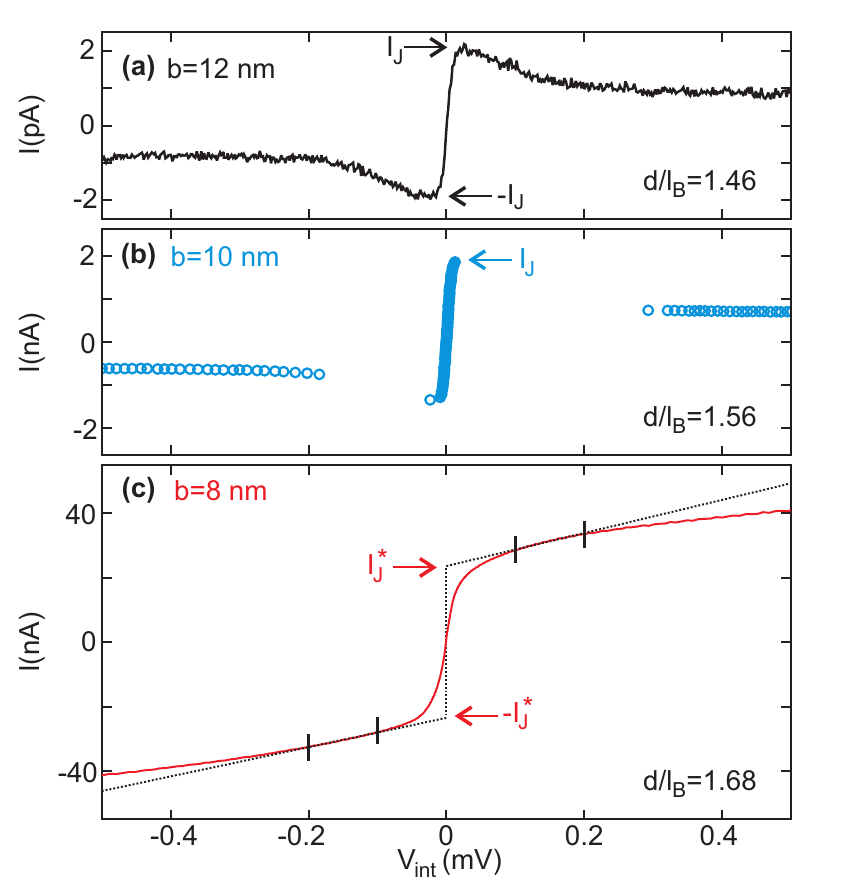}
 \centering
\caption{\label{Fig01}
(Color online)~Interlayer $I$-$V$ characteristics at $\nu_{tot}=1$  obtained from three bilayers with comparable sample sizes but different barrier thicknesses showing their effect of the Josephson currents which are signalled by the  steep current increases terminated by critical values near zero voltage.}
\end{figure}

We start with the interlayer tunneling experiments which demonstrate the strong influence of interlayer distance on the coherent phenomenon. Here the source and drain contacts are placed at the two layers separately. By sweeping the bias voltage, both the interlayer current and the four-terminal interlayer voltage are recorded.  Fig.~\ref{Fig01}(a)--(c) presents the current-voltage ($I$-$V$) characteristics from three samples with different barrier thicknesses. All three curves show a steep increase of the current around zero bias signalling the Josephson coupling between the layers. After reaching a critical current, the interlayer voltage ($V_{int}$) starts to increase rapidly. Note, that the current scales of the three plots are significantly different, while the voltage scales are the same.

Despite the overall similarity, there are significant differences in the details of the curves. For $b$=12~nm, the $I$-$V$ characteristic agrees very closely with theory~\cite{Hyart2011} and with the ones observed experimentally in devices with a similar tunneling barrier  \cite{Spiel2004}. The interlayer current reaches the nA-range when the barrier thickness is reduced from 12 to 10~nm (Fig.~\ref{Fig01}(b)). Also, the $I$-$V$ characteristics undergo a drastic change as a result of the barrier thickness reduction. Data gaps occur in regions with negative conductances and hysteresis exists in the two opposite sweep directions (data shown in Fig.~\ref{Fig01} was obtained when sweeping $V_{int}$ upward.). Essentially, the negative differential conductance in the sample with $b=10$~nm causes    the load line effect to become prominent. A further reduction of the barrier to $b$=8~nm causes the interlayer $I$-$V$ to change once more. The current is not only much larger but also the $I$-$V$ curve is smooth again and does not show interruptions. The smoothness is generic for this type of sample with $b$=8~nm. The same behavior has been observed on multiple samples from two wafers.

\begin{figure}
 \includegraphics[width=86mm]{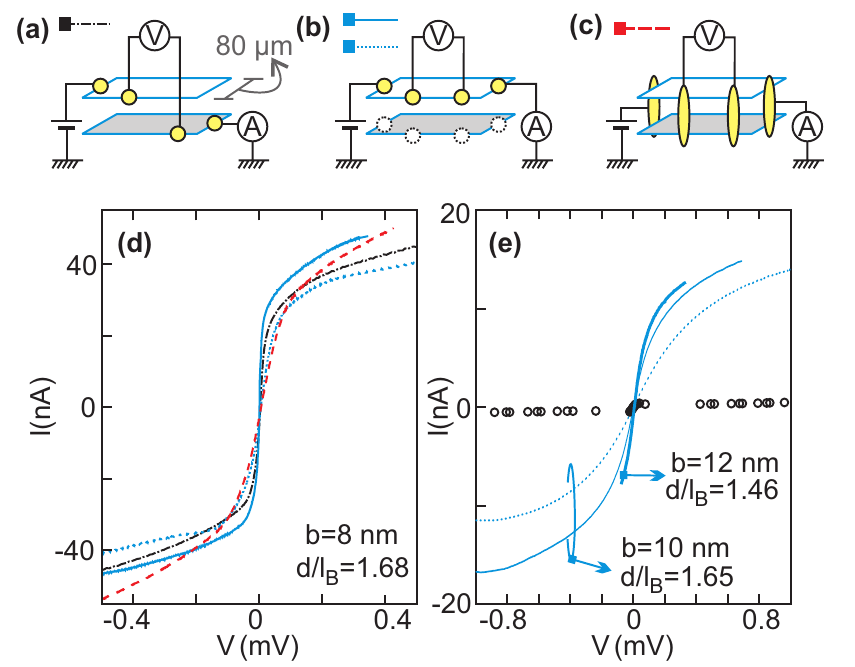}
 \centering
  \caption{\label{Fig02}(Color online)~(a) --(c) Three measurement configurations on a bilayer with a Hall bar geometry and the corresponding line styles of the data: (a) interlayer tunneling experiment: dash-dotted; (b) intralayer study on the top (bottom) layer: solid (dotted); (c)  intralayer study on both layers: dashed. (d) $I$-$V$ characteristics of the sample with $b=$8~nm at $\nu_{tot}=1$ measured using the respective configurations on the top. (e) Intralayer transport studies at $\nu_{tot}=1$ on samples with 10 and 12 nm barriers. For comparison, the interlayer $I$-$V$ characteristic obtained from the sample with $b=$10~nm is shown (circles).}
\end{figure}

The complete disappearance of negative conductance sections in the $I$-$V$ merits further investigation. We find that due to the strong Josephson coupling, the two layers in the sample with $b=8$~nm become electrically indistinguishable
such that interlayer and intralayer $I$-$V$ characteristics are identical. Note that this strong coupling only occurs at $\nu_{tot}=1$. The circuits for the interlayer and intralayer experiments on a Hall bar are sketched in panel (a) to (c) on the top of Fig.~\ref{Fig02}. Fig.~\ref{Fig02}(a) is the setup we have used for the interlayer experiment. The source and drain contacts contact the two layers separately. Furthermore, they are situated on the two opposite ends of the Hall bar. Fig.~\ref{Fig02}(b) presents a configuration where only one of the layers is exclusively selected for the measurement whereas Fig.~\ref{Fig02}(c) shows that both layers are contacted. The latter two cases reveal the intralayer transport properties.

Data obtained from the sample with $b=$8~nm using the three configurations are shown in Fig.~\ref{Fig02}(d).
Three curves marked as solid, dotted and dashed show that as soon as the in-plane current exceeds a critical current of about 20~nA the intralayer transport becomes dissipative hence the longitudinal voltage increases rapidly. This can be interpreted as a manifestation of the breakdown of the $\nu_{tot}=1$ state in a way which is very similar to the breakdown of the traditional integer quantum Hall effect due to the onset of dissipative processes \cite{Nachtwei1999}. Strikingly, the $I$-$V$ curve from the interlayer experiment (dash-dotted curve) follows closely the same breakdown behavior as the two intralayer ones.  It indicates that a Josephson current of about 20~nA produces the same dissipative effect in the intralayer transport in this geometry. Larger Josephson currents are not accessible, since the dissipative intralayer process becomes important.

With larger barriers the Josephson phenomenon can be separated from the breakdown in the intralayer transport. In Fig.~\ref{Fig02}(e) at first, the interlayer (circles) and the intralayer (lines) transport of a $b$=10 nm sample are compared. Here the critical Josephson current is about 10 times smaller than the breakdown current of the quantum Hall state. Also shown in the same panel is part of the intralayer $I$-$V$ characteristic of the sample with a 12 nm barrier. The displayed voltage interval is limited because of a breakdown of the electrical separation of the two layers outside that range. Nevertheless, it appears that the intralayer breakdown current is in the same 10 nA range as the samples with $b=$8 and 10~nm. In contrast to an exponential dependence on the barrier thickness followed by the interlayer Josephson currents, the breakdown currents exhibit only a weak barrier dependence.

\begin{figure}
 \includegraphics[width=86mm]{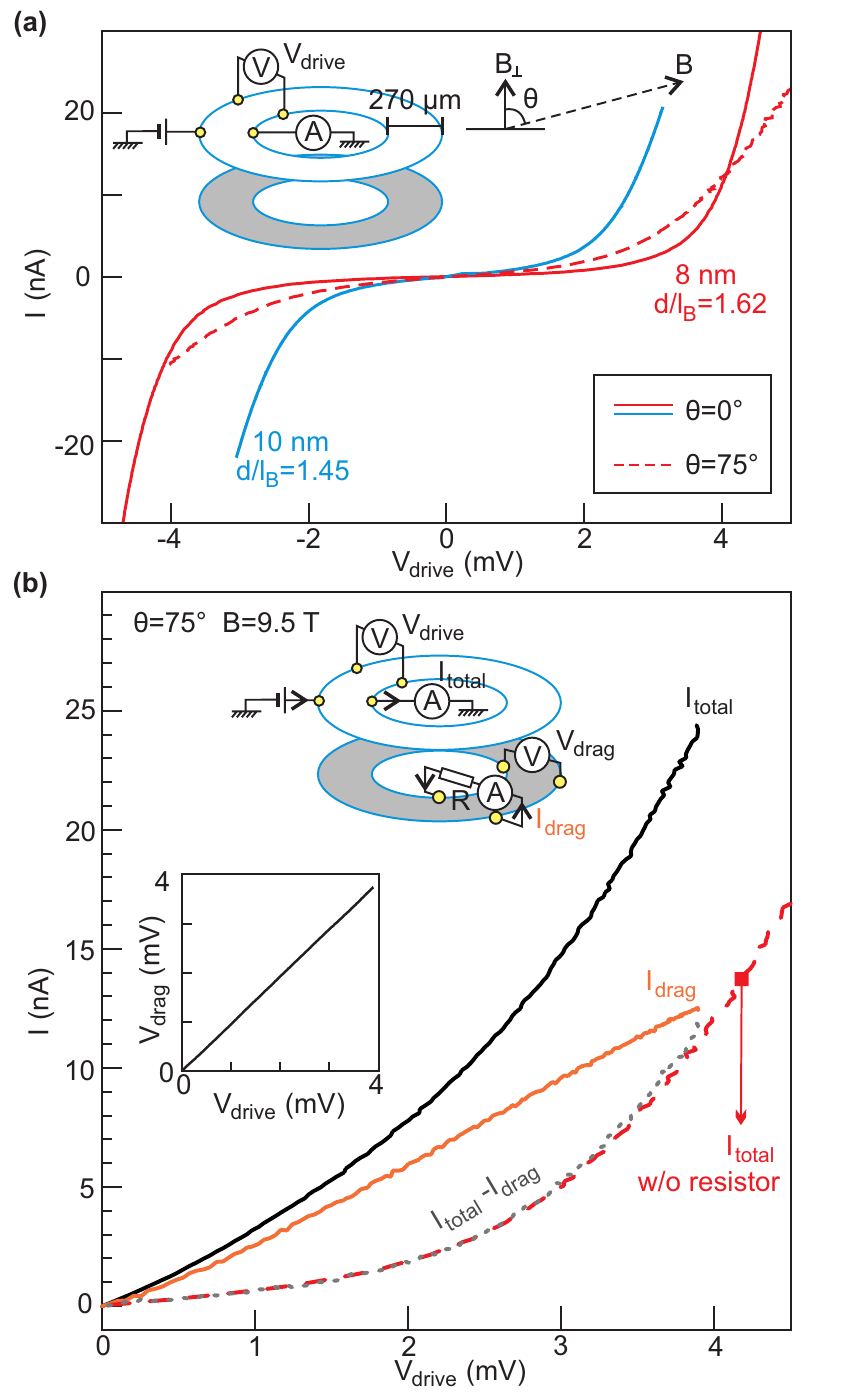}
 \centering
 \caption{\label{Fig03}
 (Color online)~(a) Intralayer $I$-$V$ characteristics at $\nu_{tot}=1$ on Corbino samples with $b=$8 and 10~nm. Curves are obtained  when contacting only the top layer. The dashed curve is measured when the Corbino sample with $b$=8~nm is rotated by 75 degrees while keeping the perpendicular $B$ field fixed. (b) $I$-$V$ characteristics obtained with the Coulomb drag configuration (top inset). Current directions are defined by the arrows. Under a tilt angle of 75 degrees, the $\nu_{tot}=1$ state is found at a total $B$-field of 9.5~T. Inset in the middle compares the drive and drag voltages recorded simultaneously.
 }
\end{figure}

The breakdown of the $\nu_{tot}=1$ state can also be observed using a Corbino geometry. The inset to Fig.~\ref{Fig03}~(a) shows the setup for a bilayer system. A (drive) voltage is applied between the outer and inner edges of the top layer. The current across the annulus is monitored as function of the interedge voltage. As the voltage increases just from zero, the current is vanishingly small. It reflects the insulating bulk property of a quantum Hall system, i.e. $\sigma_{xx}\approx0$. Above a critical voltage the current starts to increase rapidly signaling the collapse the quantum Hall state, again in close analogy with conventional quantum Hall systems.

Exploring the same Corbino samples but under a tilted $B$-field reveals that the same breakdown mechanism leads also to the failure of the Coulomb drag when the Josephson effect is suppressed by an in-plane magnetic field. Fig.~\ref{Fig03} (b) displays our results from the sample with $b$=8~nm (a sample with $b$=10 nm showed the same behavior). A large in-plane $B$-field is implemented by rotating the sample while keeping the perpendicular $B$-field fixed such that $nh/(eB_{\perp})=\nu_{tot}=1$.  The $\nu_{tot}=1$ state survives at the tilt angle of 75 degrees, but the interlayer current is below 20~pA in the interlayer voltage range -3 to 3~mV . With the interlayer current greatly suppressed, one can exclusively study the intralayer transport~\cite{Nandi2012}. We employ the Coulomb drag setup sketched on the top left panel of Fig.~\ref{Fig03}~(b). The directions of currents are defined by the arrows.

When no resistor is placed across the lower layer, the circuit is the same as the one presented in Fig.~\ref{Fig03}~(a). Likewise, the $I$-$V$ characteristic shows that the current can only flow across the annulus at large bias voltages when the $\nu_{tot}=1$ state breaks down (dashed curve in Fig.~3~(a)(b)). The finite conductance at small voltages is due to the weakened $\nu_{tot}=1$ state under tilted $B$-field. By bridging the outer and inner rims of the bottom layer with a resistor, currents start to flow in both layers but in opposite directions (top layer: outer to inner; bottom layer: inner to outer.). At small bias, the ratio between the drag current $I_{drag}$ and the total current $I_{total}$ is close to one indicating a perfect Coulomb drag. When $V_{drive}$ exceeds 2~mV, $I_{total}$ deviates from $I_{drag}$ significantly---$I_{drag}$ keeps increasing almost linearly whereas $I_{total}$ increases much faster.  Interestingly, by subtracting the drag current from the total current, the $I$-$V$ characteristic (dotted) collapses back to the dashed curve obtained when no resistor was mounted. It indicates that the excess current in $I_{total}$ originates from the breakdown of the $\nu_{tot}=1$ state. Notably, even when the $\nu_{tot}=1$ state starts to break down, the counterflow current---represented by $I_{drag}$---seems to be unaffected in the voltage range we measured (2 to 4 mV). In other words, the coherent counterflow current is still present and can flow simultaneously with the incoherent and dissipative charge current.

\begin{figure}
 \includegraphics[width=86mm]{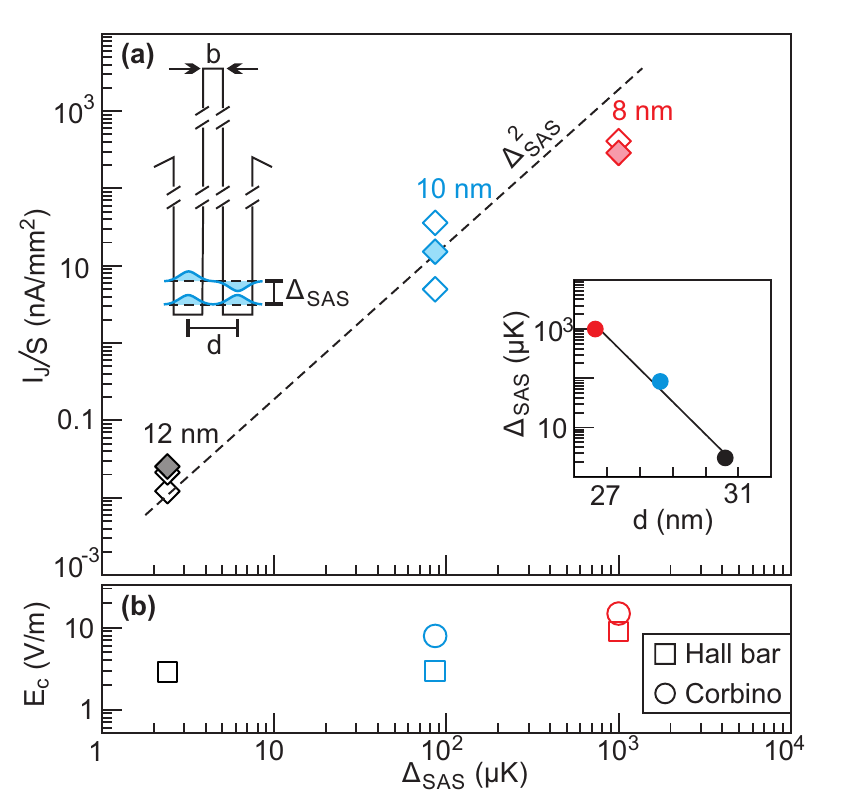}
 \centering
  \caption{\label{Fig04}(Color online)~(a)~Normalized critical current as a function of $\Delta_{SAS}$.  Filled data points are obtained from the curves shown in Fig.~\ref{Fig01}~(a) to (c). For each $\Delta_{SAS}$ value, results for several $d/l_B$ ratios are plotted: from top to bottom: $d/l_B=1.46,~1.48,~1.62$ for $b=12$~nm; $d/l_B=1.42,~1.56,~1.65$ for $b=10$~nm; $d/l_B=1.39,~1.68$ for $b=8$~nm (from two wafers). The dashed line shows the trend: $I\propto\Delta_{SAS}^2$. The left inset gives a sketch of the conduction band profile. The single particle tunneling amplitude $\Delta_{SAS}$ is defined as the energy spacing between the lowest symmetric and antisymmetric subbands. The right inset shows the dependence of the evaluated $\Delta_{SAS}$ on the interlayer distance $d$. The solid line is a linear fit. (b)~Critical Hall electric fields for the $\nu_{tot}=1$ quantum Hall state.}
\end{figure}

Finally, we discuss the values of the critical Josephson currents as well as those of the breakdown thresholds in the three batches of samples in Fig.~\ref{Fig04}.
For samples with $b$=12 and 10~nm, the critical Josephson currents ($I_J$) are well-defined (marked by the arrows in Fig.~\ref{Fig01}~(a) and (b)). For the sample with $b=8$~nm, we define $I_J$ from the crossing of a linear fit to the high voltage data ($|V|\in[0.1,0.2]$~mV) at zero bias. We use $I_J^\ast$ to remind that this is not the intrinsic Josephson critical current but the breakdown current of the $\nu_{tot}=1$ quantum Hall state. These critical current values are normalized by the sample areas ($S$) for comparison~\cite{Finck2008,Tie2009}. They are plotted as a function of $\Delta_{SAS}$--the subband splitting that measures the single particle tunneling strength. The values of $\Delta_{SAS}$ are evaluated from the interlayer conductance measured at zero $B$-field and follow the expected exponential reduction with the interlayer distance (inset to Fig.~\ref{Fig04}~(a)). This justifies our use of $\Delta_{SAS}$ as the $x$-axis of Fig.~\ref{Fig04}. For completeness, data obtained at different $d/l_B$ ratios are also included. In general, the critical Josephson current follows the quadratic dependence on $\Delta_{SAS}$. This gives a more stringent support to the perturbation theory~\cite{Hyart2011} as only one parameter---the interlayer distance---is varied here. A deviation comes from samples with $b=8$~nm reflecting that the breakdown threshold on the intralayer transport has been reached. Fig.~\ref{Fig04} also suggests that the largest intrinsic Josephson critical current that is not limited by the intralayer breakdown threshold will be reached at a $\Delta_{SAS}$ of around 500~$\mu$K.

Analyzing the breakdown behavior of the intralayer transport may shed light on the microscopic mechanism of the dissipative processes in the $\nu_{tot}=1$ state. The electric Hall fields at the respective breakdowns are summarized in  Fig.~\ref{Fig04}(b)  for both Hall bar and Corbino samples. Their values do not differ significantly between the two sample geometries and, as already discussed, do not increase much with $\Delta_{SAS}$. They are, however, about three orders of magnitude smaller than those observed at the traditional integer quantum Hall effect~\cite{Nachtwei1999}. We suggest that the breakdown at $\nu_{tot}=1$ occurs when the electric Hall field is sufficient to loosen the excited quasiparticles which are vortices with charge $e/2$, also called merons, that reside in puddles induced by disorder. The typical binding energy $\Delta_{\nu=1}$ of the merons to their respective sites is about 0.1~K as deduced from the thermal activation experiments~\cite{Wie2004}. With the observed critical Hall fields ($E_c=8\sim 15$~V/m for Corbino samples) an average distance between the merons can be estimated: $\Delta_{\nu=1}/(eE_c)=600\sim1000$~nm. Our estimated length is only slightly larger than the length scale (a few hundred nm) derived theoretically by Hyart and Rosenow~\cite{Hyart2011}.

To summarize, we have shown that the Josephson coupling in the bilayer excitonic condensate state is greatly affected by the interlayer distance or by the single particle tunnel coupling constant $\Delta_{SAS}$. It is quantitatively well described by the theory of Hyart and Rosenow \cite{Hyart2011}. Intra layer transport shows a breakdown similar to the one observed at the integer quantum Hall effect. In the $\nu_{tot}=1$ state the relevant process appears to be the loosening of merons in contrast to the inter Landau level scattering at the quantum Hall effect.
The Coulomb drag experiment on a Corbino sample under tilted $B$-field  shows that the charge current of the mobile merons coexist with the excitonic counterflow current over a wide electric field range.

We acknowledge discussions  with
J. Smet,
B. Rosenow,
T. Hyart,
A. H. MacDonald,
P. Eastham,
S. Schmult and
L.Tiemann.
M. Hagel assisted in the sample preparation.
The experimental techniques are based upon earlier collaborations with J.G.S. Lok and L.Tiemann.
This project was supported by the BMBF (German Ministry of Education and Research) Grant No 01BM900.
\bibliographystyle{apsrev}

\end{document}